\documentclass[a4paper,11pt]{article}
\usepackage{pos}
\usepackage{xspace}

\title{
Prompt neutrinos from the atmosphere to the forward region of LHC
}
\ShortTitle{
Prompt neutrinos from the atmosphere to the forward region of LHC
}

\author[a]{Weidong Bai}
\author[b]{Milind Diwan}
\author*[c]{Maria Vittoria Garzelli}
\author[d]{Yu Seon Jeong}
\author[e]{Mary Hall Reno}

\affiliation[a]{School of Physics, Sun Yat-sen University, 
Guangzhou, Guangdong 510275, P. R. China}
\affiliation[b]{Brookhaven National Laboratory, Upton, New York, USA}
\affiliation[c]{Universit\"at Hamburg, {II}. Institute for Theoretical Physics,\\
Luruper Chaussee 149, 22761 Hamburg, Germany}
\affiliation[d]{High Energy Physics Center, Chung-Ang University,
Dongjak-gu, Seoul 06974, Republic of Korea}
\affiliation[e]{Department of Physics and Astronomy, University of Iowa, Iowa City, IA 52242, USA}

\emailAdd{baiwd3@mail.sysu.edu.cn}
\emailAdd{diwan@bnl.gov}
\emailAdd{maria.vittoria.garzelli@desy.de}
\emailAdd{yusjeong@cau.ac.kr}
\emailAdd{mary-hall-reno@uiowa.edu}

\abstract{ 
  We investigate the kinematical regions that are important for producing prompt neutrinos in the atmosphere and in the forward region of the LHC, as probed by different experiments. We illustrate the results as a function of the center-of-mass nucleon-nucleon collision energies and rapidities of neutrinos and of the parent heavy-flavoured hadrons. We find overlap in
  part of the kinematic space. 
}

\FullConference{The European Physical Society Conference on High Energy Physics (EPS-HEP2023)\\
 21-25 August 2023\\
Hamburg, Germany\\}

\begin{document}
\maketitle

\section{Introduction}
\label{intro}

The present study aims at exploring the connections between prompt neutrino fluxes in the atmosphere~\cite{Bhattacharya:2016jce, Zenaiev:2019ktw} and at the Large Hadron Collider (LHC)~\cite{Bai:2020ukz,Bai:2021ira,Bai:2022jcs,Bai:2022xad}. In the atmosphere, collisions of cosmic-ray (CR) nucleons/nuclei with N, O nuclei occur, whereas at the LHC in standard collider modality $pp$ collisions are by far the most studied, but $p$Pb and PbPb runs have also taken place. Further studies have involved the use of fixed targets, such as He, Ne and Ar.

Prompt neutrinos in the case of both the LHC and the atmosphere are those that come from the production and decay of heavy-flavour hadrons, as opposite to conventional neutrinos, which come from the production and decay of light-flavour hadrons, especially $\pi^\pm$, $K^\pm$, $K^0_S$, $K^0_L$~\cite{Lipari:1993hd,Gaisser:2016uoy}. The difference comes from the different proper decay lengths of the involved hadrons, for instance the proper decay length for $D^\pm$, $c\tau_{0,D^\pm} = 0.031$ cm, is much shorter than those for charged pions and kaons $c\tau_{0,\pi^\pm}$~=~780~cm and $c\tau_{0,K^\pm}$~=~371~cm.

\begin{figure}[h!]
  \centerline{
  \includegraphics[width=0.64\textwidth]{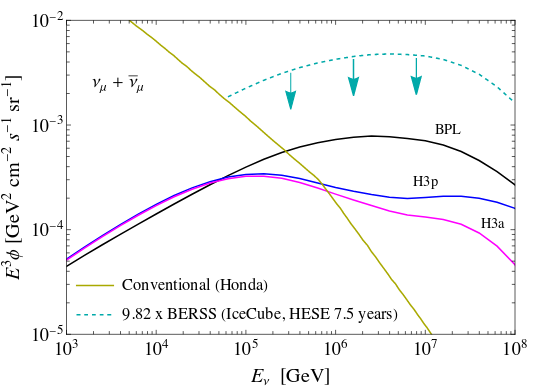}}
  \caption{\label{fig:trans} Transition from the conventional (solid yellow) to the prompt atmospheric ($\nu_\nu + \bar{\nu}_\mu$) flux, using three different models for CR spectrum composition~\cite{Gaisser:2011klf} (black: broken-power law, blue: H3p, pink: H3a).  The IceCube upper limit on prompt neutrino flxu from the analysis of 7.5 years of High-Energy  Starting Events~\cite{IceCube:2020wum} is also reported (light-blue dashed line).} 
\end{figure}

The effect of this difference, in combination with the fact that
the deepness of the atmosphere is limited, plays a role in the atmosphere. In fact one can define a critical energy $\epsilon_c$ for each kind of hadron, inversely proportional to the proper decay length and proportional to the atmospheric height, and observe that for energies above $\epsilon_c$, hadron decay probability is suppressed with respect to the probability of further interactions. The fact that the critical energy for $\pi^\pm$ and $K^\pm$ is much smaller than for heavy-flavour hadrons, leads to the fact that the most energetic atmospheric neutrinos mostly come from heavy-flavour hadrons. The prompt neutrino flux clearly overcomes the conventional one for $E_{\nu,LAB} \gtrsim 6 - 10 \cdot 10^{5}$ GeV, as shown in Fig.~\ref{fig:trans} (see also Ref.~\cite{Garzelli:2015psa, Zenaiev:2019ktw}). The exact position of the transition point also depends on the composition of CRs, on which many uncertainties still exist at energies above the so-called ``knee''~\cite{AlvesBatista:2019tlv, Albrecht:2021cxw}.

On the other hand, one does not see this effect for neutrinos measured in far-forward experiments at the LHC, 
where however the presence of elements along the beam line (like e.g. magnets), absent in the atmosphere, can deviate charged particles non decaying promptly and thus affect the rates of neutrinos at detectors. At present, two far-forward experiments exist, Faser$\nu$~\cite{FASER:2019dxq,FASER:2022hcn} and SND@LHC~\cite{SHIP:2020sos, SNDLHC:2022ihg}, taking data in Run 3 at the LHC, located at $\sim$ 480 m from the ATLAS interaction point, on the tangent to the accelerator arc. In any case, at far-forward experiments at the LHC, it will be possible to distinguish $\nu_\tau$ and $\bar{\nu}_\tau$ fluxes~\cite{Bai:2020ukz, Bai:2021ira, Bai:2022jcs}, that are exclusively of prompt origin. When looking at the $\nu_e$
and $\bar{\nu}_e$ fluxes at the LHC, one expects that their high-energy tail is still dominated by prompt neutrinos. On the other hand, the $\nu_\mu$ and $\bar{\nu}_\mu$ fluxes are largely dominated by conventional neutrinos, even at high energies, in contrast to the atmospheric case of Fig.~\ref{fig:trans}.

\section{Framework for theoretical calculations}
\begin{figure}
\begin{center}
  \includegraphics[width=0.49\textwidth]{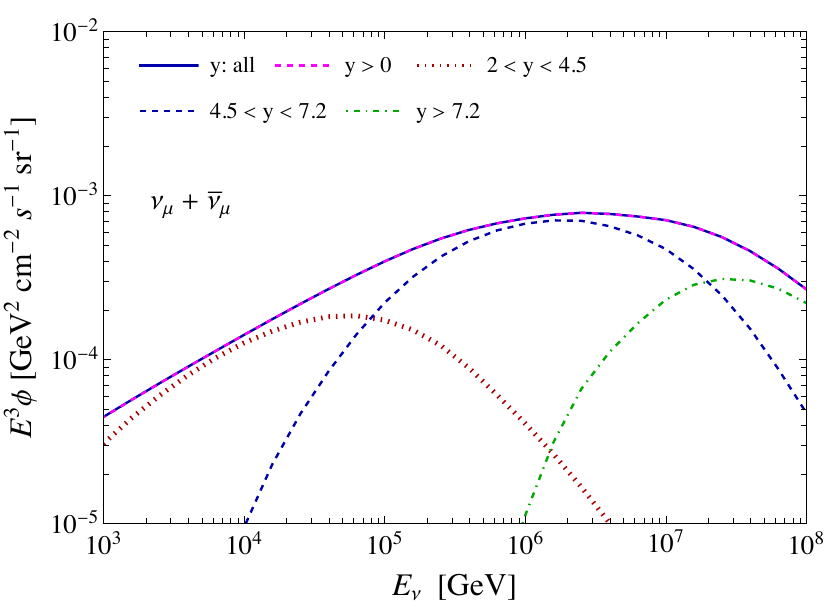}
  \includegraphics[width=0.49\textwidth]{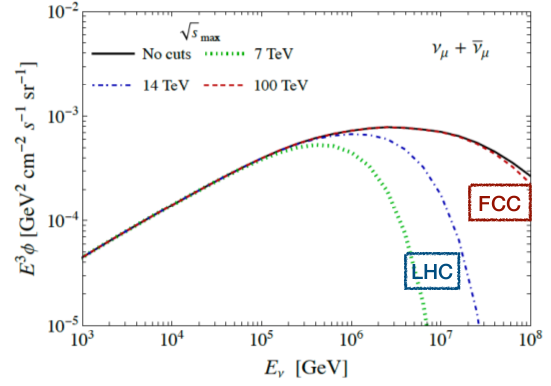}
  \caption{\label{fig:decompo}
    Prompt $(\nu_\mu + \bar{\nu}_\mu)$ flux: contribution of neutrinos and neutrinos from charmed hadrons in different rapidity ranges (left panel); contribution of neutrinos and antineutrinos from nucleon-nucleon collisions with COM energies $\sqrt{s}$ up to 7, 14 and 100 TeV, corresponding to Run 1 at the LHC, HL-LHC and FCC-hh, respectively (right panel).
 See Ref.~\cite{Bai:2022xad} for more detail.}
\end{center}
\end{figure}

To establish a connection between prompt neutrinos in the atmosphere and at the LHC, we first observe that different regions of the atmospheric prompt neutrino spectrum are populated by neutrinos produced by the decay of charmed hadrons in different rapidity ranges, as shown in the left panel of Fig.~\ref{fig:decompo}.
In particular, as outlined in Ref.~\cite{Bai:2022xad}, ($\nu_\mu +\bar{\nu}_\mu$) at the $E_{\nu, LAB} \sim \mathcal{O}$(PeV) scale come mostly from $D$-mesons in the rapidity range $4.5 < y < 7.2$. This interval is above but immediately adjacent to the one explored by studies of $D$-meson production at LHCb, $2 < y < 4.5$~\cite{LHCb:2013xam, LHCb:2015swx, LHCb:2016ikn}. This basically means that, on the one hand, a collider experiment capable of directly seeing $D$-meson production in the range $4.5 < y < 7.2$, absent at the moment, would be useful to establish a connection with atmospheric prompt neutrinos. On the other hand, we expect that a theory description working well in the region where LHCb data are present, could go on working at least decently even in the rapidity range relevant for prompt atmospheric neutrinos at the PeV scale.

Looking at the right panel of Fig.~\ref{fig:decompo}, the possibility to establish connections becomes more complicated for prompt neutrinos at energies $E_{\nu, LAB} >$ 5 PeV, considering that the latter are produced by collisions at energies larger than those reachable at the LHC, typically explorable only by a future hadron-hadron collider with higher center-of-mass (COM) energy. In particular, the production of prompt neutrinos with $E_{\nu,LAB} \sim$ 100 PeV could be explored at the Future Circular Collider (FCC) at $\sqrt{s} = 100$ TeV~\cite{FCC:2018vvp,Benedikt:2022wvj}.

\begin{figure}
  \begin{center}
\begin{center}
  \includegraphics[width=0.49\textwidth]{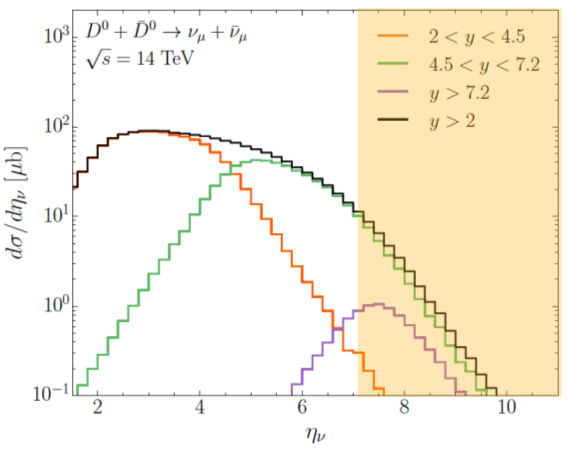}
  \includegraphics[width=0.49\textwidth]{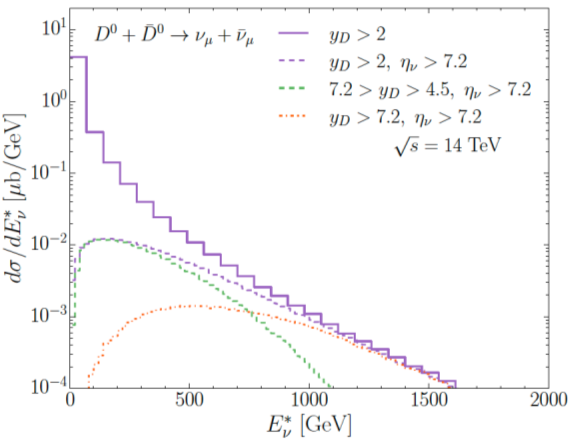}
\end{center}
    \caption{\label{fig:confor}
      Prompt $(\nu_\mu + \bar{\nu}_\nu)$ from ($D^0+\bar{D}^0$) production and decay at the LHC at $\sqrt{s}=14$ TeV: pseudorapidity spectrum (left panel) and COM energy spectrum (right panel), considering the contribution from charmed mesons in different rapidity ranges.}
\end{center}
\end{figure}

On the other hand, charmed mesons, e.g. $D^0 + \bar{D}^0$, with $4.5 < y_c < 7.2$ at the LHC give rise to $\nu + \bar{\nu}$ populating a wide rapidity spectrum, with a maximum around $\eta_\nu \sim 5$, as shown in the left panel of Fig.~\ref{fig:confor}, and a tail slowly decreasing with increasing rapidity. In particular, it turns out that the majority of prompt neutrinos populating the region $\eta_\nu > 7$ comes from charmed mesons in the region $4.5 < y_c < 7.2$, because charmed meson production in the region $y_c > 7$ is very much suppressed. Although the latter is still dominant when looking at the highest energy tail of the energy spectrum of neutrinos with $\eta_\nu > 7$, the low energy tail of the spectrum corresponding to the bulk of the fiducial cross-section, is mostly populated by neutrinos from charmed hadrons with $4.5 < y_c < 7.2$. This is well visible in the right panel of Fig.~\ref{fig:confor}, where it is clear that the neutrinos from charmed mesons with $y_c > 7.2$ overcome those from charmed mesons with $4.5 < y_c < 7.2$ only above a neutrino energy of $\sim$ 700 GeV, in the COM frame.

It follows that present and future far-forward neutrino experiments~\cite{Anchordoqui:2021ghd}, although
being sensitive only to very-narrow very-forward pseudorapidity ranges, i.e. 
to the region $\eta_\nu \gtrsim 7$ or $\eta_\nu \gtrsim 8$, still probe the
wider kinematic region for charm production which is relevant for prompt neutrino fluxes in the atmosphere.
It will be then enough to focus on the low energy part, up to a few
hundred GeV, of the neutrino spectra measured by far-forward neutrino experiments. At that point, the issue, still to be overcome, will be the fact that, at such low energies, conventional neutrinos will still dominate both the $\nu_\mu + \bar{\nu}_\mu$ and $\nu_e + \bar{\nu}_e$ LHC fluxes. The $\nu_\tau$ and $\bar{\nu}_\tau$ spectra will then remain the most useful for establishing a direct connection between measurements at the LHC and in the atmosphere.

\section{Conclusions}
\label{sec:conclu}

Prompt neutrinos in the atmosphere at energies $E_{\nu, LAB} \sim \mathcal{O}$(PeV), as relevant for recent IceCube studies and as a background for searches of neutrinos from far astophysical sources, are mainly produced by collisions at energies explorable at the LHC, giving rise to charmed mesons in the range $4.5 < y_c < 7.2$, undergoing (semi)leptonic decays after production. Present and future far-forward neutrino experiments at the LHC measuring neutrinos with $\eta_\nu \gtrsim 7$ or $\eta_\nu \gtrsim 8$, are also sensitive to the same charmed-meson rapidity region, considering that neutrinos from those mesons abundantly populate the neutrino energy spectrum under $\eta_\nu \gtrsim 7$ or $\eta_\nu \gtrsim 8$ cuts and dominate its low energy part. This establishes a connection between prompt neutrinos in the fields of accelerator and astroparticle physics.
Considering that, however, the low energy part of the $\nu_e + \bar{\nu}_e$ and $\nu_\mu + \bar{\nu}_\mu$ energy spectra at the LHC is also heavily populated/dominated by conventional neutrinos, measurements of $\nu_\tau$ and $\bar{\nu}_\tau$ at the LHC, all of prompt origin, will be particularly useful for exploiting the aforementioned connection. 
The highest-energy tail of the prompt atmospheric neutrino flux, corresponding to $E_{\nu, LAB} \sim 10^7 - 10^8$~GeV, however, could be directly explored only at a future more energetic collider, like FCC, as follows from the fact that the highest-energy tail of the cosmic-ray spectrum (corresponding to the so-called ankle region and beyond it) extends to energies much higher than the LHC ones. 
An extensive discussion on these topics and more detail can be found in Ref.~\cite{Bai:2022xad}. 

\section*{Acknowledgements}
This work has been supported in part by U.S. Department of Energy Grant DE-SC-0010113 and DE-SC-0012704 and the National Research Foundation of Korea (NRF) grant funded by the Korea government Ministry of Science and ICT (MSIT)
No. 2021R1A2C1009296. The work of M.V.G. has been supported in part by the Bundesministerium f\"ur Bildung und Forschung under contract 05H21GUCCA.

\bibliographystyle{JHEP}
\bibliography{nuproc}

\end{document}